\documentclass[journal]{IEEEtran}

    \def\BibTeX{{\rm B\kern-.05em{\sc i\kern-.025em b}\kern-.08em
        T\kern-.1667em\lower.7ex\hbox{E}\kern-.125emX}}
    \usepackage[backend=biber, style=ieee, sorting=none]{biblatex}
    \usepackage{graphicx}
    \ifCLASSINFOpdf
    \else
    \fi
    \usepackage{amsmath}
    \usepackage{amssymb}
    \makeatletter
    \newcommand{\eqpunct}[1]{%
    \gdef\@eqpunct{#1}%
    }
    \gdef\@eqpunct{}

    \def\tagform@#1{%
    \maketag@@@{(\ignorespaces#1\unskip)\@eqpunct}%
    \gdef\@eqpunct{}%
    }
    \makeatother

    \usepackage{algorithm}
    \usepackage{algorithmic}

    \usepackage[dvipsnames]{xcolor}
    \definecolor{myred}{HTML}{D20F39}
    \definecolor{myblue}{HTML}{1E66F5}
    \definecolor{mygreen}{HTML}{40A02B}

    \usepackage[colorlinks = true, linkcolor = black, citecolor = black, urlcolor = black]{hyperref}
    \usepackage{orcidlink}

    \usepackage{siunitx}

\begin{document}
    %
    \title{Peregrino: A Full-Hardware Accelerator for the Complete Falcon Post-Quantum Digital Signature Scheme on Resource-Constrained Edge Devices}
%
%
%

\author{Antonio Carreño, Jaime Señor \orcidlink{0000-0001-7695-9152}, and Jorge Portilla \orcidlink{0000-0003-4896-6229}\thanks{Antonio Carreño, Jaime Señor (corresponding author), and Jorge Portilla are with the Centro de Electrónica Industrial y Sistemas Multimodales, Universidad Politécnica de Madrid, 28006 Madrid, Spain (e-mail: a.cgomez@upm.es; jaime.senors@upm.es; jorge.portilla@upm.es).}
}

%
%

\markboth{Carreño \MakeLowercase{\textit{et al.}}: Full-Hardware Accelerator for Falcon}%
{Carreño \MakeLowercase{\textit{et al.}}: Full-Hardware Accelerator for Falcon}
%



\maketitle

\begin{abstract}
The arrival of quantum computers threatens the security guarantees of classical cryptography, since quantum algorithms can break schemes that remain secure against conventional attacks. The National Institute of Standards and Technology (NIST) has therefore standardized a set of post-quantum cryptographic algorithms, among them Falcon, a lattice-based digital signature scheme with the most compact signature and public-key sizes of the standardized candidates. Falcon's reliance on floating-point arithmetic makes it hard to implement in hardware, and prior work offers only partial accelerators for specific operations such as signature generation or verification, or a single full implementation generated through high-level synthesis (HLS). This work presents Peregrino, the first hardware accelerator of the complete Falcon digital signature scheme designed from scratch in HDL, targeting resource-constrained edge devices without a native floating-point unit through an emulated floating-point datapath. Implemented on a single Artix 7 XC7A200T FPGA, Peregrino uses 85261 LUTs, 41382 FFs, 44 BRAMs, and 142 DSPs for the Falcon-1024 variant. Against the only prior full implementation, the HLS-based FalconTakesOff, it uses 1,9$\times$ fewer LUTs, 3,4$\times$ fewer FFs, 2,7$\times$ fewer BRAMs, and 9,9$\times$ fewer DSPs, fitting the entire scheme on one FPGA where the HLS design requires at least two. Operating as a peripheral of an on-chip MicroBlaze soft-core, the accelerator additionally reduces key-pair generation, signature generation, and signature verification clock cycles by 92\%, 96\%, and 85\% over the emulated floating-point reference software.
\end{abstract}



%
\IEEEpeerreviewmaketitle

\section{Introduction}\label{sec:introduction}
%
%
%
%
\IEEEPARstart{T}{he} arrival of quantum computers threatens the security of classic cryptographic schemes which are mostly used today. In particular, they are able to perform efficient large integer factorization via Shor's algorithm \cite{Shor}, which is theoretically able to break classic public-key cryptographic schemes like RSA or Elliptic Curve Cryptography (ECC) in much less time than current supercomputers do \cite{gidney2025factor2048bitrsa}.

This new quantum computer era makes it necessary to develop new cryptographic schemes which protect information from both algorithms used by current supercomputers and algorithms used by quantum computers. This is the reason why the National Institute of Standards and Technology (NIST) opened a post-quantum cryptography (PQC) standardization process in 2016 \cite{NIST2016}. In September 2022, in the third round of the post-quantum cryptography standardization process \cite{NIST3Round}, four schemes were selected for their standardization: one Key Encapsulation Mechanism (KEM), that is CRYSTALS-Kyber \cite{CRYSTALSKyber}, and three Digital Signature Algorithms (DSA), which are CRYSTALS-Dilithium \cite{CRYSTALSDilithium}, Falcon \cite{FALCON}, and SPHINCS+ \cite{SPHINCSplus}. Afterwards, an additional fourth round \cite{NIST4Round} was performed to select another KEM whose security is not based on the same mathematical problems as CRYSTALS-Kyber. In this regard, the HQC scheme \cite{HQC} was chosen for its standardization. The standards for both Falcon and HQC are not yet officially public, but the ones for CRYSTALS-Kyber, CRYSTALS-Dilithium, and SPHINCS+ can be found in Federal Information Processing Standards (FIPS) 203 \cite{FIPS203}, FIPS 204 \cite{FIPS204}, and FIPS 205 \cite{FIPS205}, respectively.

Fast Fourier Lattice-based Compact Signatures Over NTRU, better known as Falcon, is a PQC DSA based on three fundamental principles: NTRU lattices, GPV theoretical framework \cite{GPV}, and an optimized sampler using the Fast Fourier Transform (FFT). This algorithm was selected due to its signature and key compactness, and the execution speed of the signature verification among others. However, this scheme has two features which make its hardware design more challenging to implement than the other digital signature schemes. First feature is the complex and recursive structure in certain parts that leads to non-trivial comprehension. Second feature is the use of floating-point arithmetic, which for hardware platforms or processors that do not have a Floating-Point Unit (FPU) implies an additional difficulty. These are the reasons why currently there is less background about hardware design of Falcon in comparison with the other digital signature schemes, as well as the lack of hardware implementations designed from scratch.

In order to solve the problem of floating-point numbers, the Hawk \cite{HAWK} lattice-based DSA was developed. This scheme achieves greater performance than Falcon in key-pair and signature generation while getting rid of the floating-point arithmetic. On the other hand, it increases the size of public keys and reduces performance in signature verification, two reasons why Falcon was selected by NIST. This trade-off between performances and compactness still makes Falcon worth researching.

The goal of this work is to design Peregrino, the first hardware accelerator
made from scratch of the entire Falcon digital signature scheme (key-pair
generation, signature generation, and signature verification), equivalent
to the official reference software implementation \cite{FALCONurl}. Unlike
the only other complete hardware implementation, which is generated through
high-level synthesis, a hand-written design gives full control over the
datapath, which this work uses to achieve substantially lower resource
utilization. It is intended to make the most of the advantages offered by
Hardware Description Language (HDL) design such as parallelism and modular
organization to develop a more optimized version of the scheme. The
contributions of this work are listed below:

\begin{enumerate}
    \item Designing a hardware accelerator made from scratch of the entire Falcon digital signature scheme (key-pair generation, signature generation, and signature verification), capable of operating as a peripheral for CPUs and aiming to operate on the edge in resource-constrained environments.

    \item Implementing a modular design in which every block can be replaced with an equivalent block chosen by the user. This fact will increase the portability and adaptability of the accelerator for different environments and applications.

    \item Using a memory centric architecture to solve the problems related to shared memory access by several hardware modules.

    \item Implementing this design in a commercial FPGA (\textit{Xilinx XC7A200T-1SBG484C}) and validating its proper functioning, as well as carrying out an investigation about the resource utilization and performance.

    \item Demonstrating that a hand-written RTL implementation brings the complete scheme within the resource budget of a single mid-range FPGA: Peregrino uses 1,9$\times$ fewer LUTs, 3,4$\times$ fewer FFs, 2,7$\times$ fewer BRAMs, and 9,9$\times$ fewer DSPs than the only prior full implementation, whose combined resource demand exceeds the capacity of one device of this class.
\end{enumerate}

The rest of the paper is organized as follows: Section \ref{sec:relatedwork} presents a brief review about the partial hardware accelerators of Falcon digital signature scheme available in the literature, Section \ref{sec:falconpreliminaries} contains a theoretical base of the behavior of the scheme and its algorithms, Section \ref{sec:HardwareDesign} explains which design decisions have been taken related to the adaptation to hardware design, and Section \ref{sec:results} shows the results associated with the test bench carried out after implementation on the target FPGA in comparison to the software reference and the state of the art. Finally, in Section \ref{sec:discussion} a discussion about the implementation performance is followed by a final conclusion in Section \ref{sec:conclusion}.

\section{Related Work}\label{sec:relatedwork}

Once NIST selected the multiple PQC schemes for their standardization, the research community has developed numerous hardware accelerators with the aim to achieve greater performance than the software reference. In this section, existing Falcon implementations on FPGAs are analyzed to provide a baseline for comparison with the proposed design.

In the case of Falcon, there is not a large number of hardware implementations on FPGA due to a couple of fundamental problems. The first problem is the recursive and complex structure of the algorithm, which hinders comprehension and makes it difficult to implement it in hardware. The second is the use of floating-point numbers, an additional problem for hardware designs as they do not have native FPUs.

To the best of the authors' knowledge, the only design developed in hardware of the entire Falcon scheme is an implementation synthesized from software code using High Level Synthesis (HLS) \cite{FalconTakesOff}. In this implementation the authors use the pragmas offered by HLS design such as array partitioning, loop unrolling, function inlining, pipelining or dataflow to optimize the algorithm. The authors also reformulate the recursive functions of the tree computation to make them synthesizable to HDL.

The hardware/software codesign implementations are very varied in content, from designs which run one of the three main parts of the algorithm such as signature verification to designs optimizing basic operation blocks. One example of the first type of implementations is FalconSign \cite{FalconSign}, where signature generation is performed with reduced latency aiming at real-world implementations. A memory centric architecture with an FPU is used, in addition to a sampler modification to reduce latency and increase throughput. Another implementation that covers a main part of the algorithm is the one made by Beckwith et al. \cite{Beckwith}. In this work an accelerator for the entire CRYSTALS-Dilithium scheme and the signature verification of Falcon is developed. They optimize the verification by running decoding and message hashing in parallel hardware, which is efficient depending on the size of the message.

In addition, a recent hardware/software co-design targets an Arty-7 board \cite{nguyen}, but key-pair generation runs entirely in software, and only the pseudo-random number generator, a modular adder, and the verification block are implemented on the FPGA. Likewise, \cite{shrivastava} integrates custom instructions into a 32-bit Vex RISC-V core to accelerate signature generation and verification, reporting speedups of approximately 1,6$\times$ and 4,3$\times$ respectively over a software baseline on the same core. EFX \cite{lee_efficient_2024} instead targets granular, operation-level optimization across Falcon's functions to conserve silicon area rather than offloading a single state, reporting a 3,58$times$ reduction in signing cycles over a prior hardware accelerator through a co-design that runs the sampling function concurrently in software and hardware.

The second type of implementations goes deeper into specific function of the Falcon algorithm. The implementation presented in \cite{BiSamplerZ} benefits from the sequential double execution of the SamplerZ function as it uses independent data and it can be parallelized. The authors consider that both sequential functions are not simultaneously in the same stage, so the module is pipelined to be more efficient in case of rejection, and an assistance mechanism is designed so that the successful datapath helps the rejected datapath to recompute intermediate values. Recently in \cite{OutrunningMillenium} another implementation is designed to optimize SamplerZ function. In this design the authors develop three different versions in which circuit-level and architecture-level innovations are combined such as sampling parallelization without increasing the resource utilization. Algorithm modifications are carried out, such as changing the Horner sequential method to the Estrin scheme that groups coefficients recursively to evaluate them in parallel.

In \cite{RNS} authors focus on the NTRUSolve function part of the key-pair generation. This function makes a decomposition into the residue number system (RNS). Four modules are developed which gather both the conversion from integer to RNS, the conversion from RNS to integer, big- and small-integer multiplication, and Montgomery multiplication. These modules are supported by circular shift registers (CSR) for input and output data.

The most time- and resource-consuming parts of the Falcon algorithm are the Number Theoretic Transform (NTT), the Fast Fourier Transform (FFT), and their inverse. In \cite{CompactFFTNTT} an accelerator for both transforms is designed. It reuses basic operation modules to share them between the transforms as a processing unit.

In the case of the implementation in \cite{LightWeightFFT} a hardware accelerator of the FFT is designed. The main improvement of this design is the change from the Cooley-Tukey model to the Winograd model. This model reduces the number of multiplications necessary to perform the FFT. Additionally, it has at its disposal a reconfigurable architecture that permits varying the size of the polynomials and modifying the accelerator with an instruction set.

Finally, recently in December 2025 in \cite{EMINEM} a general NTT accelerator was designed for Falcon, CRYSTALS-Dilithium, and Hawk DSAs. This implementation takes advantage of the computation speed of the NTT Radix-4. A module using both Radix-2 and Radix-4 is designed to avoid reconfiguration depending on the polynomial degree. When the polynomial degree is not a power of four, a first round with Radix-2 is applied, and then the polynomial is divided into two parts on which the Radix-4 NTT can be applied.

\section{Falcon Preliminaries}\label{sec:falconpreliminaries}

In order to understand the decisions made at design time, it is necessary to comprehend the problems related to certain incompatibilities between software design and hardware design, and how the algorithm is structured.

Digital signature schemes are intended to protect the authenticity of the transmitted information and, by extension the content itself. Digital signatures make it possible to know if the message has been modified by third parties in the transmission and, at the same time, confirm the sender's authorship of the information. The digital signature process consists of three different parts: key-pair generation, signature generation, and signature verification.

There are three digital signature schemes selected by NIST for their standardization, of which Falcon presents the most compact signature and public key size. Compactness of Falcon makes it ideal for resource constrained environments or specific applications where this reduced size is required.

Falcon is based on three fundamental principles:

\begin{enumerate}
    \item Gentry-Peikert-Vaikuntanathan framework \cite{GPV} describes how to obtain hash-and-sign lattice-based signature schemes, and has been proven in classic random oracle models and quantum random oracle models.
    \item NTRU lattices defined in \cite{NTRU} allow to reduce the public key size in $O(n)$. This lattices are associated with the equation
\begin{equation}
    h \equiv g \cdot f^{-1} \pmod{q} 
    \label{eq:clavepublica}
\end{equation}
    for the computation of the public key and the NTRU equation
\begin{equation}
    fG-gF \equiv q \pmod{\phi}
    \label{eq:ecuacionNTRU}
\end{equation}
    for the private key.
    \item Chosen trapdoor sampler is a random variant of the ``fast Fourier nearest plane'' \cite{FastFourierNearestPlane} which allows to work with NTRU lattices in a fast and compact manner.
\end{enumerate}

Key-pair generation is marked by the solution of the NTRU equations as commented previously, as well as the generation of a tree-type structure in FFT representation. This key-pair generation is the most computationally expensive, and it is reusable for various signatures. This is the reason why some values are stored to perform multiple signatures with only a pair of keys. This pair of keys will be valid for at least several signatures in the hundreds of thousands or until a security failure is anticipated. Algorithm \ref{alg:keygen} shows how polynomials are created, the generation of the tree $\mathsf{T}$, and finally the derivation of both keys.

\begin{algorithm}[t]
\caption{\textbf{Keygen} ($\phi$,$q$)}
\label{alg:keygen}
\begin{algorithmic}[1]
\REQUIRE A monic polynomial $\phi\in{\mathbb{Z}[x]}$, a modulus $q$
\ENSURE A secret key $\mathsf{sk}$, a public key $\mathsf{pk}$
\STATE $f,g,F,G\leftarrow \operatorname{NTRUGen}(\phi,q)$
\STATE $\mathbf{B}\leftarrow\begin{bmatrix}
g & -f\\
G & -F
\end{bmatrix}$
\STATE $\hat{\mathbf{B}}\leftarrow \operatorname{FFT}(\mathbf{B})$
\STATE $\mathbf{G} \leftarrow \hat{\mathbf{B}} \times \hat{\mathbf{B}}^*$
\STATE $\mathsf{T} \leftarrow \operatorname{ffLDL}^*(\mathbf{G})$
\FOR{each leaf $\mathsf{leaf}$ of $\mathsf{T}$}
\STATE $\mathsf{leaf.value} \leftarrow\sigma/\sqrt{\mathsf{leaf.value}}$
\ENDFOR
\STATE $\mathsf{sk} \leftarrow(\hat{\mathbf{B}}, \mathsf{T})$
\STATE $h\leftarrow gf^{-1}\pmod{q}$
\STATE $\mathsf{pk} \leftarrow h$
\RETURN $\mathsf{sk}, \mathsf{pk}$
\end{algorithmic}
\vspace{2pt}
\hrule
\vspace{2pt}
\centering \textit{Falcon key-pair generation algorithm \cite{FALCON}.}
\end{algorithm}

Signature generation includes both the hash of the message desired for transmission and the use of the fast Fourier sampler. The ultimate goal is to obtain a signature that satisfies the equation

\begin{equation}
    s_1+s_2h \equiv c \pmod{q}\ .
    \label{eq:ecuacionfirma}
\end{equation}

Algorithm \ref{alg:sign} presents the signature generation process. The sampler function will be shown in more detail later on since, due to its recursive structure additional modules are designed to adapt it to hardware design.

\begin{algorithm}[t]
\caption{\textbf{Sign} ($\mathsf{m}, \mathsf{sk}, \lfloor\beta^2\rfloor$)}
\label{alg:sign}
\begin{algorithmic}[1]
\REQUIRE A message $\mathsf{m}$, a secret key $\mathsf{sk}$, a bound $\lfloor\beta^2\rfloor$
\ENSURE A signature $\mathsf{sig}$ of $\mathsf{m}$
\STATE $\mathsf{r} \leftarrow\{0,1\}^{320}$ uniformly
\STATE $c\leftarrow \operatorname{HashToPoint}(\mathsf{r}\mid\mid \mathsf{m},q,n)$
\STATE $\mathbf{t}\leftarrow(-\frac{1}{q}\operatorname{FFT}(c)\odot \operatorname{FFT}(F),\frac{1}{q}\operatorname{FFT}(c)\odot \operatorname{FFT}(f))$
\REPEAT
\REPEAT
\STATE $\mathbf{z}\leftarrow \operatorname{ffSampling}_n(\mathbf{t},\mathsf{T})$
\STATE $\mathbf{s} = (\mathbf{t} - \mathbf{z}) \hat{\mathbf{B}}$
\UNTIL $\|\mathbf{s}\|^2\leq \lfloor\beta^2\rfloor$
\STATE $(s_1,s_2)\leftarrow \operatorname{invFFT}(s)$
\STATE $\mathsf{s}\leftarrow \operatorname{Compress}(s_2,8\cdot \mathsf{sbytelen}-328)$
\UNTIL $(\mathsf{s}\neq\perp)$
\RETURN $\mathsf{sig} = (\mathsf{r}, \mathsf{s})$
\end{algorithmic}
\vspace{2pt}
\hrule
\vspace{2pt}
\centering \textit{Falcon signature generation algorithm \cite{FALCON}.}
\end{algorithm}

Signature verification is a fast and simple process in comparison with key and signature generation. In this case the receiver hashes the salt and message to a lattice point $c$, uses the public key to recover the second signature component, and accepts only if the resulting short vector's norm falls below a pre-established bound. Algorithm \ref{alg:verify} shows this process, while its details can be found in \cite{FALCON}.

\begin{algorithm}[t]
\caption{\textbf{Verify} ($\mathsf{m}, \mathsf{sig}, \mathsf{pk}, \lfloor\beta^2\rfloor$)}
\label{alg:verify}
\begin{algorithmic}[1]
\REQUIRE A message $\mathsf{m}$, a signature $\mathsf{sig} = (\mathsf{r}, \mathsf{s})$, a public key $\mathsf{pk} = h \in \mathbb{Z}_q[x]/(\phi)$, a bound $\lfloor\beta^2\rfloor$
\ENSURE Accept or reject
\STATE $c \leftarrow \operatorname{HashToPoint}(\mathsf{r}\mid\mid \mathsf{m},q,n)$
\STATE $s_2 \leftarrow \operatorname{Decompress}(\mathsf{s},8\cdot \mathsf{sbytelen}-328)$
\IF{$s_2 = \perp$}
\STATE reject
\ENDIF
\STATE $s_1 \leftarrow c -s_2h\pmod{q}$
\IF{$\|(s_1,s_2)\|^2 \leq \lfloor\beta^2\rfloor$}
\STATE accept
\ELSE
\STATE reject
\ENDIF
\end{algorithmic}
\vspace{2pt}
\hrule
\vspace{2pt}
\centering \textit{Falcon signature verification algorithm \cite{FALCON}.}
\end{algorithm}

\section{Hardware Design}\label{sec:HardwareDesign}

As this is the first hardware implementation of the entire Falcon digital signature scheme designed from scratch to the authors' knowledge, Falcon software code \cite{FALCONurl} is used as reference, and from this code hardware modules are designed. The main objective of this work is to provide a functional hardware implementation of the complete algorithm, and from that baseline make optimizations to either increase performance or reduce the resource utilization. Development from scratch enables to have complete control of the hardware accelerator final design, and to have the capability of modifying the code to achieve an identical functionality to the reference code. This feature is essential for trying to take full advantage of parallelism offered by hardware design and increase the number of operations per clock cycle, reducing the execution time. Nevertheless, coming from a sequential programming language like C to a concurrent programming language like VHDL generates several incompatibilities which must be solved for the proper design of the accelerator. These problems are listed below:

\begin{enumerate}
    \item Software functions are reused without generating additional resources.
    \item Software code uses pointers and memory allocation accessible from every function in the scheme.
    \item Data from several zones of the memory are used with different types depending on the section of the algorithm that is being executed and the required data sizes.
    \item Floating-point types are used in this scheme, that are not native to the hardware programming language.
    \item Multiplications of large numbers are performed, leading to high logic delay in hardware implementation.
    \item There are recursive functions which store the internal state of the program and do not generate additional resources.
\end{enumerate}

As a result of these problems, a series of design decisions have been made to adapt software code to hardware. These design decisions are explained in the following sections.

\subsection{Modular Design}\label{sec:ModularDesign}

\begin{figure}[t]
    \centering
    \includegraphics[width = \columnwidth]{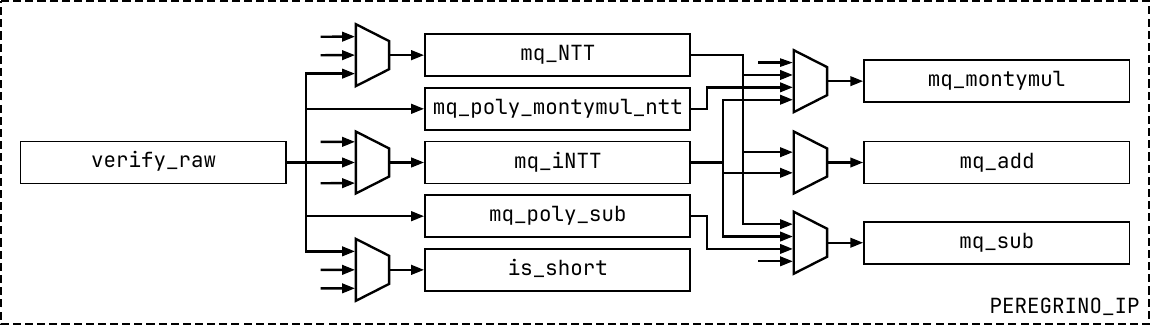}
    \caption{Tree-like modular design.}
    \label{fig:ModularDesign}
\end{figure}

The implementation presented in this work has a modular structure. In this structure every module can be replaced with another equivalent module chosen by the developer. Each module corresponds to a software function of the reference code, making them reusable in different parts of the scheme. These modules are deployed with a structure similar to the Fig. \ref{fig:ModularDesign}, in which modules communicate between them under request. This communication combined with the use of multiplexers solves the problem of reusing modules. Two upper modules can request operations from a lower one without doubling hardware resources as long as this requests are not made simultaneously.

\begin{figure}[t]
    \centering
    \includegraphics[width = \columnwidth]{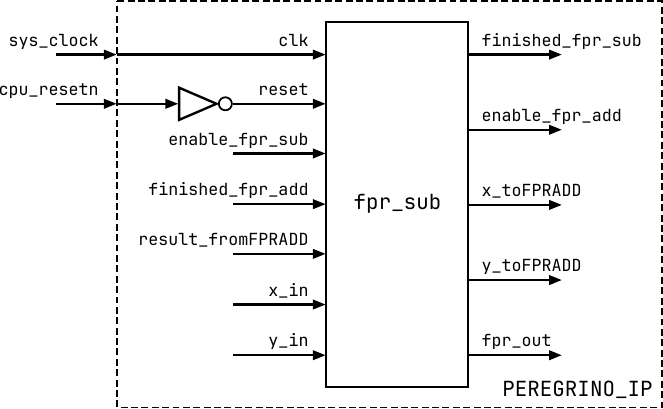}
    \caption{fpr\_sub module example.}
    \label{fig:Module}
\end{figure}

Every module has an interface type shown in Fig. \ref{fig:Module} with the example of \textit{fpr\_sub}. Enable signals from upper modules can be observed, as well as control signals to lower modules, and data transmission signals between modules or memories. This makes any developer implement their specific module modifying the interface to communicate with this design, making this work portable and flexible.

\subsection{Memory Centric Architecture}\label{sec:MemoryCentric}

\begin{figure}[t]
    \centering
    \includegraphics[width = \columnwidth]{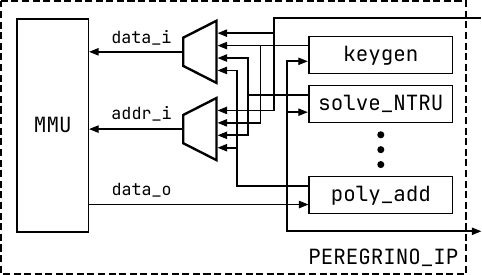}
    \caption{Memory-centric architecture.}
    \label{fig:MemoryCentric}
\end{figure}

Reference code uses pointers to access memory positions from several functions of the scheme. This implies an obstacle in hardware development as no conventional memory exists. To overcome this problem the hardware design includes BRAMs to store data temporarily. These BRAMs are usually for the exclusive use of one or a couple of modules, leading to numerous data transmission stages between modules. It has been decided that a memory centric architecture will be implemented in this design, being accessible from any module by signals which emulate the pointer method. Memory-centric architecture is seen in Fig. \ref{fig:MemoryCentric}.

\begin{figure*}[t]
    \centering
    \includegraphics[width = .8\textwidth]{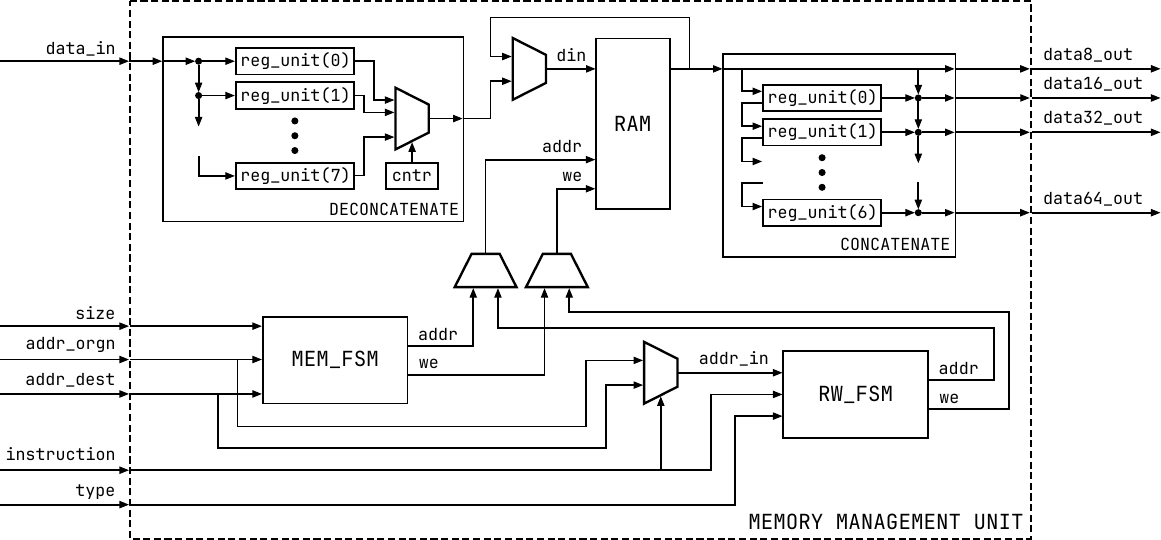}
    \caption{Memory management unit.}
    \label{fig:MMU}
\end{figure*}

This provides a central memory that is easily accessible from all modules, increasing similarities between hardware modules and software functions when designing hardware descriptions.

\subsection{Memory Management Unit}\label{sec:MMU}

One of the main problems in the algorithm is the use of different data types as explained in the previous sections. A central memory, together with a memory management unit (MMU) capable of handling 8, 16, 32, and 64 bits of data, solves this problem. The MMU consists of five modules: the central memory, a finite state machine (FSM) for reading and writing, another FSM for memory movements, a segmentation module before the memory, and a concatenation module after the memory. The structure of this MMU is shown in Fig. \ref{fig:MMU}.

The central memory has enough space to store the keys, the Falcon tree computations, and temporary buffers used in Falcon-1024. As several data types are used, 8-bit size has been selected as base word length because it is the minimum size among the data types used in the algorithm. Reading and writing FSM receives a signal to select the type of operation, a signal to indicate the data size, and a memory address. This FSM is responsible for managing memory addresses corresponding to bigger data sizes, as a 32-bit word will need to access 4 memory positions. Memory movement FSM receives the zone of the memory transferred expressed in bytes, an origin address, and a destination address. These two FSMs allow to send needed signals from the modules, enable the process, and wait until the operation is finished. Concatenation module stores data coming from the central memory in shift registers and provides data with its proper size combining the registers values. Lastly, segmentation module receives a 64-bit signal, divides it into bytes, and stores the values in registers to send them to the central memory.

\subsection{Floating Point Arithmetic}\label{sec:FloatingPoint}

The usage of floating-point data types implies an additional difficulty in environments without native FPU such as hardware design or low-end processors. Falcon authors defined different directives to operate with data in emulated floating-point as described in IEEE-754 standard \cite{IEEE754}. This standard defines 64-bit data divided into sign, exponent, and mantissa.

\begin{itemize}
    \item Most significant bit corresponds to the sign. Its value will be 0 when data is positive and 1 otherwise.
    \item Next 11 bits correspond to exponent, from bit 52 to bit 62. This part stores the power of two the mantissa has to be multiplied with.
    \item Remaining 52 bits, from bit 0 to bit 51, correspond to mantissa. This part stores the number's magnitude.
\end{itemize}

Software functions using this emulated floating-point directive have been selected as a reference for designing the hardware modules. To learn how to manage values that may generate conflict or how operations are performed in detail, consult the reference code.

\subsection{Karatsuba Algorithm}\label{sec:KaratsubaModule}
Floating-point data previously mentioned in subsection \ref{sec:FloatingPoint}, among others, are stored in 64-bit signals and operate with them. Operations such as addition, subtraction or bit shifting do not lead to problems. However, multiplications between large data types are particularly difficult to implement in hardware. Multiplications implemented in hardware require applying certain number of resources which generate some delay as they use several logic units. During synthesis process these multiplications are introduced into digital signal processing (DSP) units. The \textit{Xilinx XC7A200T-1SBG484C} FPGA has DSPs with 25x18 two's complement multiplier with 48-bit accumulator, which means for one large number multiplication up to four DSPs are needed. The operation will be done in just one clock cycle, but the concatenation of several DSPs increases the logic delay and therefore the final clock frequency of the implementation is reduced.

The solution adopted to this problem has been to isolate the large number multiplication and apply the Karatsuba algorithm \cite{karatsuba1963} to them to reduce the size of the multiplications. A systematic comparison of Baseline, Tiling, Comba, and Karatsuba multiplier designs for Falcon on FPGA and ASIC targets is given in \cite{magesh_multiplers}, which supports the choice of Karatsuba adopted here. This algorithm is based on the idea that a multiplication can be reduced by splitting its operands in half, multiplying smaller numbers, and adding intermediate operations. In turn these smaller multiplications could be reduced by applying recursively the Karatsuba algorithm. To implement this algorithm in hardware some modifications are needed to operate with binary format instead of decimal format, this modification is shown in Algorithm \ref{alg:karatsuba}. Bit shifting operations with powers-of-two values are used to replace multiplications with powers-of-ten values. Two accelerators have been designed, one of them uses Karatsuba algorithm in 64-bit multiplications and in some sensitive 32-bit multiplications, and the other accelerator only applies Karatsuba in 64-bit multiplications. Comparison between these two accelerators will be performed, one of them operating at greater frequency but lasting more clock cycles, to analyze which one is the most efficient option.

\begin{algorithm}[t]
\caption{\textbf{Binary Karatsuba Algorithm} ($x$,$y$)}
\label{alg:karatsuba}
\begin{algorithmic}[1]
\REQUIRE Two operands $x$, and $y$
\ENSURE A product $z$
\STATE $x_h \leftarrow x(\mathsf{size}-1$ $\text{downto}$ $\mathsf{size}/2)$
\STATE $x_l \leftarrow x(\mathsf{size}/2 - 1$ $\text{downto}$ $0)$
\STATE $y_h \leftarrow y(\mathsf{size}-1$ $\text{downto}$ $\mathsf{size}/2)$
\STATE $y_l \leftarrow y(\mathsf{size}/2 - 1$ $\text{downto}$ $0)$
\STATE $p_h\leftarrow x_h \cdot y_h$
\STATE $p_l\leftarrow x_l \cdot y_l$
\STATE $s_x\leftarrow x_h + x_l$
\STATE $s_y\leftarrow y_h + y_l$
\STATE $p_m \leftarrow s_x \cdot s_y$
\STATE $s_m \leftarrow p_m - p_h - p_l$
\STATE $z \leftarrow (p_h \ \texttt{<<}\  \mathsf{size}) + (s_m \ \texttt{<<}\ \mathsf{size}/2) + p_l$
\end{algorithmic}
\end{algorithm}

\subsection{\textit{ffSampling} Recursive Function}\label{sec:ffSampling}

The usage of recursive functions is a problem in hardware environments, and efficient execution in terms of time would involve replicating modules as many times as the function is used. Nevertheless, this approach leads to the implementation of vast resource utilization which is not available in the FPGA. That is why the division of this function into two associated modules has been proposed. One of these modules oversees doing the operations of the algorithm and requesting lower modules execution. The second module stores the intermediate states of the algorithm and also controls the state flow of the actuation module.

\begin{figure}[t]
    \centering
    \includegraphics[width = \columnwidth]{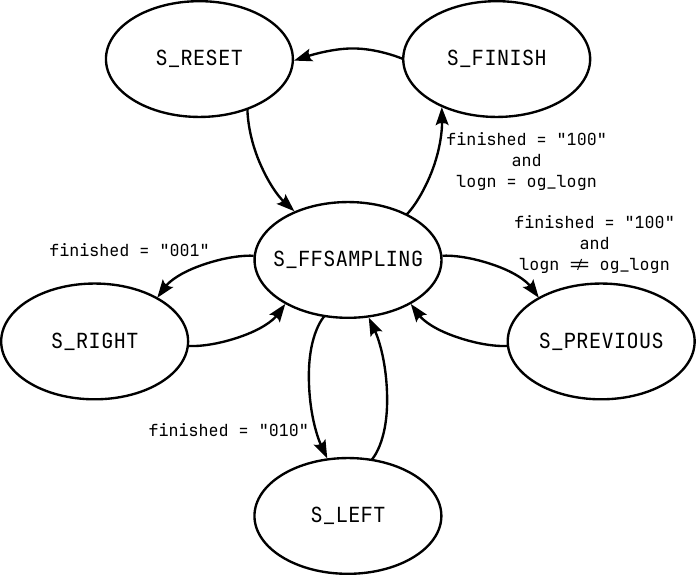}
    \caption{FSM\_Sampling state diagram.}
    \label{fig:FSMSampling}
\end{figure}

State control module distinguishes between requesting execution from the actuation module, ascending to the `father' state, and relegating to the left or right `child' following the flow diagram in Fig. \ref{fig:FSMSampling}. In states \textit{S\_RIGHT} and \textit{S\_LEFT} values of the memory addresses, polynomial degree, and state checkpoint are stored in a custom stack to recover after execution. Finally, \textit{S\_FFSAMPLING} state enables actuation module using auxiliary signals to jump into the corresponding part of the function: before the first recursive call, between the first and the second recursive calls or after the second recursive call to finalize.

Actuation module follows the software reference code behavior but it is divided into three subroutines splitting operations by recursive calls of the operation. It is the state control module that is in charge of indicating the actuation module which subroutine must follow depending on the part of the tree the operation is located at the moment and the values of the custom stack.

\subsection{Accelerator IP Interface}\label{sec:IPInterface}

In addition to a wrapper that groups every module in the architecture, an interface is needed for the communication between the accelerator and processor. This interface must be able to transmit the following instructions when requested by the processor:

\begin{itemize}
    \item Writing a 48 character seed in RAM for the pseudo-random context generation.
    \item Reading values from the pseudo-random context to complete the codified digital signature.
    \item Writing the message to be signed or the message to be verified in RAM.
    \item Reading and writing public key and digital signature values in the MMU.
    \item Enabling the computation of the key-pair generation, signature generation, and signature verification.
\end{itemize}

Data transmission between the processor and accelerator has been unified aiming to reduce the number of registers. Data size and memory addresses are handled internally to avoid synthesis errors. These data transfers are controlled with read, write, and ready signals, which allow software to manage different processes of the accelerator. Signal \textit{ram\_select} distinguishes between writing in the RAM for seed storage and message storage. Lastly, control of the execution of the different parts of the scheme consists of two signals \textit{IP\_start} and \textit{IP\_ready} to enable computation, and to check if accelerator finished execution. There is also another 3-bit signal \textit{acc\_select} that selects among key-pair generation, signature generation, and signature verification. Value in \textit{logn\_in} signal is used to select the polynomial degree and signal \textit{error\_bit} indicates if verification has been successful.

\section{Implementation Results}\label{sec:results}

The hardware accelerator design presented in this work has been implemented in an Artix 7 FPGA (\textit{Xilinx XC7A200T-1SBG484C}) that does not have a processing system. This is the reason why a MicroBlaze has been included in the programmable logic of the FPGA to act as a soft-core. The tests consisted of running the algorithm with 1000 seed-message pairs measuring the number of clock cycles along execution, since certain parts of the algorithm are susceptible to varying in time depending on the input seed. Number of clock cycles have been measured using the AXI Timer IP from Xilinx, and they have been stored using the AXI UART Lite IP communicating with an external program in the host computer. Furthermore, all reported results are for the Falcon-1024 variant.

Prior to perfomance characterization, Peregrino's functional correctness was validated at three levels. First, the intermediate outputs of each module were compared against the reference software implementation. Second, signatures produced by Peregrino verify correctly under the reference software, and signatures produced by the reference software verify correctly under Peregrino's hardware verifier. Third, the official NIST KAT vectors were run through the design and reproduced exactly.

Vitis IDE has been used to modify the software reference code and program the MicroBlaze processor. Software reference code using emulated floating point has been tested as well as using a native FPU for floating-point arithmetic, the results of this test will be the starting point to compare the accelerator performance. The number of cycles is measured also using the accelerator as a peripheral of the MicroBlaze modifying the reference code with the functions generated by default as hardware drivers of the platform. These tests are performed with both the version applying Karatsuba algorithm in 64-bit multiplications and the version applying Karatsuba algorithm also in 32-bit multiplications. Finally, some tests run the software reference code on the Pynq-Z2's ARM Cortex-A9 processor.

\begin{figure*}[t]
    \centering
    \includegraphics[width = \textwidth]{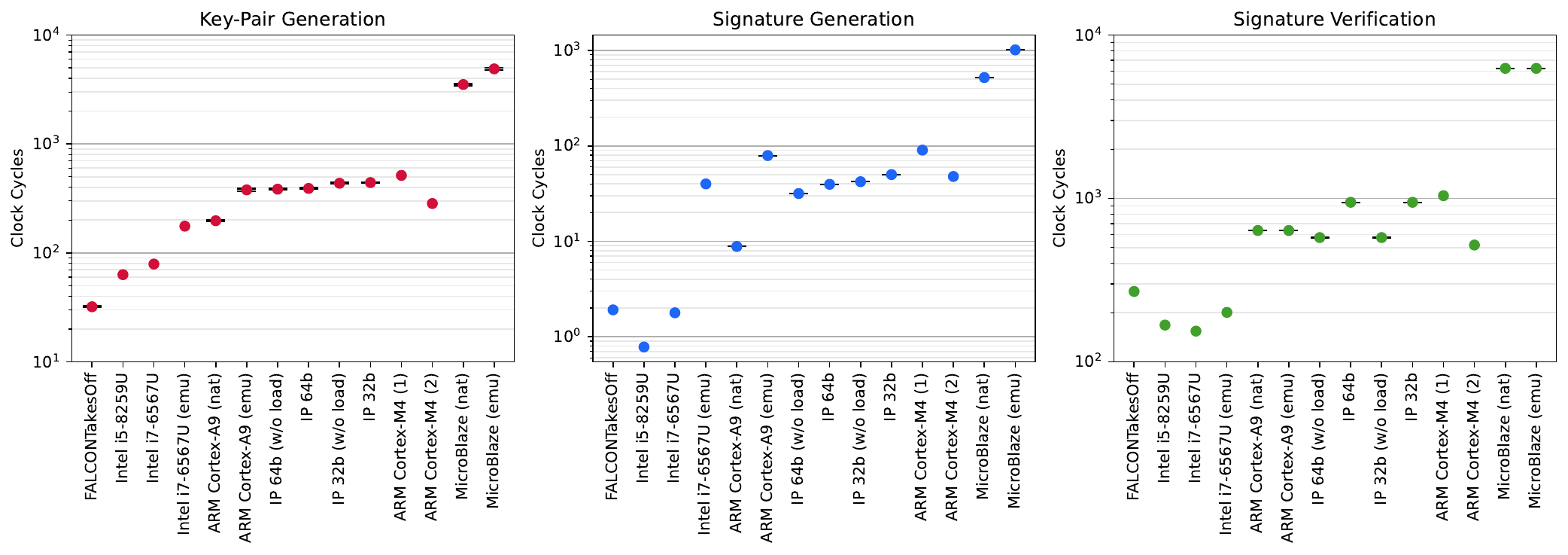}
    \caption{Mean value (with 95\% confidence interval) of clock cycles for different platforms.}
    \label{fig:Results}
\end{figure*}

Software reference code on which this work has been developed follows the directive \textit{FALCON\_FPEMU} for processors and environments without native FPU. Each test consists of using a different seed-message pair to run the entire algorithm (key-pair generation, signature generation, and signature verification). In the case of processors with native FPU, software code follows the directive \textit{FALCON\_FPNATIVE} that enables native floating-point types. In Tables \ref{tabla:compKG}, \ref{tabla:compSD}, and \ref{tabla:compVV} a comparison between the tests made and the state of the art is shown for key-pair generation, signature generation and signature verification respectively. Fig. \ref{fig:Results} shows the mean value of the number of clock cycles with a 95\% confidence interval (for those cases in which data is available), since parts of the algorithm are dependent on the input, such as key generation or the version of hash functions implemented in this work.

\begin{table*}[p]
\centering
\caption{Performance comparison of key-pair generation in Falcon-1024.\\Peregrino 64-bit (w/o load) implementation is used as the base reference for performance ($1 \times$).}
\begin{tabular}{|c|r|r|r|}
\hline
\textbf{Implementation} & \textbf{Frequency} (\si{\mega\hertz}) & \textbf{Clock Cycles} (Performance) & \textbf{Latency} (\si{\milli\second}) \\ \hline\hline
FalconTakesOff \cite{FalconTakesOff} & 100 & 32.030.000 (\textcolor{myred}{12,01}$\times$) & 320,30 \\ \hline
Intel i5-8259U \cite{FALCON} & 2300 & 63.135.000 (\textcolor{myred}{6,09}$\times$) & 27,45 \\ \hline
Intel i7-6567U \cite{Pornin2019} & 3600 & 79.164.000 (\textcolor{myred}{4,86}$\times$) & 21,99 \\ \hline
Intel i7-6567U (emu) \cite{Pornin2019} & 3600 & 168.624.000 (\textcolor{myred}{2,28}$\times$) & 46,84 \\ \hline
ARM Cortex-A9 (nat) & 650 & 197.267.098 (\textcolor{myred}{1,95}$\times$) & 303,49 \\ \hline
ARM Cortex-M4 \cite{Pornin2025} & 24 & 284.031.714 (\textcolor{myred}{1,35}$\times$) & 11.834,65 \\ \hline
ARM Cortex-A9 (emu) & 650 & 378.898.746 (\textcolor{myred}{1,01}$\times$) & 582,92 \\ \hline
Peregrino 64-bit (w/o load)  & 75 & 384.528.461 (1,00$\times$) & 5.127,05 \\ \hline
Peregrino 64-bit & 75 & 390.715.456 (\textcolor{mygreen}{0,98}$\times$) & 5.209,54 \\ \hline
Peregrino 32-bit (w/o load) & 80 & 435.899.844 (\textcolor{mygreen}{0,88}$\times$) & 5.448,75 \\ \hline
Peregrino 32-bit & 80 & 442.086.839 (\textcolor{mygreen}{0,87}$\times$) & 5.526,09 \\ \hline
ARM Cortex-M4 \cite{Pornin2019} & 168 & 513.950.073 (\textcolor{mygreen}{0,75}$\times$) & 3.059,23 \\ \hline
MicroBlaze (nat) & 100 & 3.513.814.725 (\textcolor{mygreen}{0,11}$\times$) & 35.138,15 \\ \hline
MicroBlaze (emu) & 80 & 4.890.909.836 (\textcolor{mygreen}{0,08}$\times$) & 61.136,37 \\ \hline
\end{tabular}
\label{tabla:compKG}
\end{table*}

\begin{table*}[p]
\centering
\caption{Performance comparison of signature generation in Falcon-1024.\\Peregrino 64-bit (w/o load) implementation is used as the base reference for performance comparison ($1 \times$).}
\begin{tabular}{|c|r|r|r|}
\hline
\textbf{Implementation} & \textbf{Frequency} (\si{\mega\hertz}) & \textbf{Clock Cycles} (Performance) & \textbf{Latency} (\si{\milli\second}) \\ \hline\hline
Intel i5-8259U \cite{FALCON} & 2300 & 789.564 (\textcolor{myred}{40,09}$\times$) & 0,34 \\ \hline
FalconTakesOff \cite{FalconTakesOff} & 100 & 1.638.253 (\textcolor{myred}{19,32}$\times$) & 16,38 \\ \hline
Intel i7-6567U \cite{Pornin2019} & 3600 & 1.926.252 (\textcolor{myred}{16,43}$\times$) & 0,54 \\ \hline
ARM Cortex-A9 (nat) & 650 & 8.823.999 (\textcolor{myred}{3,59}$\times$) & 13,58 \\ \hline
Peregrino 64-bit (w/o load)  & 75 & 31.649.982 (1,00$\times$) & 422,00 \\ \hline
Peregrino 64-bit & 75 & 39.541.445 (\textcolor{mygreen}{0,80}$\times$) & 527,22 \\ \hline
Intel i7-6567U (emu) \cite{Pornin2019} & 3600 & 40.297.932 (\textcolor{mygreen}{0,79}$\times$) & 11,19 \\ \hline
Peregrino 32-bit (w/o load) & 80 & 42.152.137 (\textcolor{mygreen}{0,75}$\times$) & 526,90 \\ \hline
ARM Cortex-M4 \cite{Pornin2025} & 24 & 47.778.012 (\textcolor{mygreen}{0,66}$\times$) & 1990,75 \\ \hline
Peregrino 32-bit & 80 & 50.043.600 (\textcolor{mygreen}{0,63}$\times$) & 625,55 \\ \hline
ARM Cortex-A9 (emu) & 650 & 79.199.695 (\textcolor{mygreen}{0,40}$\times$) & 121,85 \\ \hline
ARM Cortex-M4 \cite{Pornin2019} & 168 & 93.985.426 (\textcolor{mygreen}{0,34}$\times$) & 559,44 \\ \hline
MicroBlaze (nat) & 100 & 520.361.840 (\textcolor{mygreen}{0,06}$\times$) & 5.203,62 \\ \hline
MicroBlaze (emu) & 80 & 1.014.754.858 (\textcolor{mygreen}{0,03}$\times$) & 12.684,44 \\ \hline
\end{tabular}
\label{tabla:compSD}
\end{table*}

\begin{table*}[p]
\centering
\caption{Performance comparison of signature verification in Falcon-1024.\\Peregrino 64-bit (w/o load) implementation is used as the base reference for performance comparison ($1 \times$).}
\begin{tabular}{|c|r|r|r|}
\hline
\textbf{Implementation} & \textbf{Frequency} (\si{\mega\hertz}) & \textbf{Clock Cycles} (Performance) & \textbf{Latency} (\si{\milli\second}) \\ \hline\hline
Intel i7-6567U \cite{Pornin2019} & 3600 & 160.596 (\textcolor{myred}{3,59}$\times$) & 0,04 \\ \hline
Intel i5-8259U \cite{FALCON} & 2300 & 168.498 (\textcolor{myred}{3,42}$\times$) & 0,07 \\ \hline
Intel i7-6567U (emu) \cite{Pornin2019} & 3600 & 201.960 (\textcolor{myred}{2,85}$\times$) & 0,06 \\ \hline
FalconTakesOff \cite{FalconTakesOff} & 100 & 269.608 (\textcolor{myred}{2,14}$\times$) & 2,70 \\ \hline
ARM Cortex-M4 \cite{Pornin2025} & 24 & 518.594 (\textcolor{myred}{1,11}$\times$) & 21,61 \\ \hline
Peregrino 64-bit (w/o load)  & 75 & 576.009 (1,00$\times$) & 7,68 \\ \hline
Peregrino 32-bit (w/o load) & 80 & 576.009 (1,00$\times$) & 7,20 \\ \hline
ARM Cortex-A9 (emu) & 650 & 636.493 (\textcolor{mygreen}{0,90}$\times$) & 0,98 \\ \hline
ARM Cortex-A9 (nat) & 650 & 636.744 (\textcolor{mygreen}{0,90}$\times$) & 0,98 \\ \hline
Peregrino 32-bit & 80 & 947.187 (\textcolor{mygreen}{0,61}$\times$) & 11,84 \\ \hline
Peregrino 64-bit & 75 & 947.187 (\textcolor{mygreen}{0,61}$\times$) & 12,63 \\ \hline
ARM Cortex-M4 \cite{Pornin2019} & 168 & 1.032.261 (\textcolor{mygreen}{0,56}$\times$) & 6,14 \\ \hline
MicroBlaze (nat) & 100 & 6.252.922 (\textcolor{mygreen}{0,09}$\times$) & 62,53 \\ \hline
MicroBlaze (emu) & 80 & 6.252.922 (\textcolor{mygreen}{0,09}$\times$) & 78,16 \\ \hline
\end{tabular}
\label{tabla:compVV}
\end{table*}

\begin{table*}[t]
\centering
\caption{Resource utilization comparison (entire Falcon-1024 scheme).\\Peregrino 64-bit implementation is used as the base reference ($1 \times$).}
\begin{tabular}{|c|r|r|r|r|}
\hline
\textbf{Implementation} & \textbf{LUTs} & \textbf{FFs} & \textbf{BRAMs} & \textbf{DSPs} \\ \hline\hline
Peregrino 32-bit & 85750 (\textcolor{mygreen}{1,01}$\times$) & 41834 (\textcolor{mygreen}{1,01}$\times$) & 44 (1,00$\times$) & 65 (\textcolor{myred}{0,46}$\times$) \\ \hline
Peregrino 64-bit & 85261 (1,00$\times$) & 41382 (1,00$\times$) & 44 (1,00$\times$) & 142 (1,00$\times$) \\ \hline
FalconTakesOff \cite{FalconTakesOff} & 159174 (\textcolor{mygreen}{1,87}$\times$) & 141018 (\textcolor{mygreen}{3,41}$\times$) & 120 (\textcolor{mygreen}{2,73}$\times$) & 1412 (\textcolor{mygreen}{9,94}$\times$) \\ \hline
\end{tabular}
\label{tabla:computil}
\end{table*}

It should be noted that the metrics related to Peregrino with the name \textit{w/o load} represent the measures discarding the time consumption of data transfers between the CPU and the accelerator, i.e., they take into account the functioning hardware accelerator only. As it can be checked, these data transfers may consume up to 39\% of the total clock cycle count in the worst case, which corresponds to signature verification. Unless stated otherwise, the reductions reported are for the accelerator alone, while the corresponding value including MicroBlaze register-interface transfers is given alongside it where the two differ substantially.

Tests performed on the target FPGA using the MicroBlaze soft-core processor reveal that using the accelerator as a peripheral leads to a clock cycle reduction of 92\%/97\%/91\% for the accelerator alone, and 92\%/96\%/85\% once the MicroBlaze register-interface transfer cycles are included, when the processor uses emulated floating-point. Meanwhile, with a native FPU the corresponding reductions are 89\%/94\%/91\% and 89\%/92\%/85\%. These reductions are for key-pair generation, signature generation, and signature verification respectively. Key-pair generation is essentially unaffected by the distinction, since its own data transfers cost only 1,6\% of its total cycles.

When running the reference software code on an ARM Cortex-A9 processor of the Pynq-Z2 with emulated floating-point data types, there is a clock cycle reduction of 60\% in signature generation, and 10\% in signature verification while 1,5\% clock cycle increase in key-pair generation. When native FPU is used on this ARM Cortex-A9, as floating-point variables are present in key-pair generation and signature generation, a clock cycle reduction of 10\% is still achieved in signature verification.

This work is also compared with the software implementations presented by the authors in \cite{FALCON, Pornin2019, Pornin2025} over Intel i5-8259U, Intel i7-6567U and ARM Cortex-M4 processors, and with the hardware implementation designed from HLS presented in \cite{FalconTakesOff}. Reference software code implementations on Intel i5-8259U \cite{FALCON} and Intel i7-6567U \cite{Pornin2019} processors follow optimized directives such as Single Instruction Multiple Data (SIMD) to operate with vectors (AVX2), and instructions combining multiplications and additions (FMA). Nonetheless, in \cite{Pornin2019} they use one of these processors with emulated floating-point directives and the accelerator achieves a clock cycle reduction of 21\% in signature generation. Regarding the metrics from the HLS-based design in \cite{FalconTakesOff}, the clock frequencies differ from one function to another, so the latencies provided in Tables \ref{tabla:compKG}, \ref{tabla:compSD}, and \ref{tabla:compVV} consider the worst case (\SI{100}{\mega\hertz}) as if the whole scheme were deployed entirely.

Peregrino improves performance of the ARM Cortex-M4 processors in 25\%/66\%/44\% compared to the implementation made by the authors in \cite{Pornin2019} even though software code in these processors is optimized by modifying some functions at assembler level. Recently in \cite{Pornin2025}, the number of clock cycles is reduced on the ARM Cortex-M4 implementation but this work still reduces the number of clock cycles in signature generation by 34\%.

Finally, since the HLS-based design from \cite{FalconTakesOff} is the only prior work that implements the entire scheme in hardware, we provide a comparison of the resource utilization between such implementation and Peregrino in Table \ref{tabla:computil}. It should be highlighted that these numbers refer to the entire scheme (key-pair and signature generation, and signature verification). Since the work in \cite{FalconTakesOff} provides these metrics for different functions that conform the whole scheme, the presented comparison provides the sum of them.

Peregrino requires 1,9$\times$ fewer LUTs, 3,4$\times$ fewer FFs, 2,7$\times$
fewer BRAMs, and 9,9$\times$ fewer DSPs than this HLS-based design
(Table \ref{tabla:computil}).

\section{Discussion}\label{sec:discussion}

In this section we discuss the reported results and set out the decisions they justify. The main finding is that the Peregrino hardware trades peak throughput for resource efficiency: by implementing the complete Falcon scheme from scratch in HDL, it brings the full PQC DSA within the resource limitations of a single FPGA. We first examine the trade-offs of the implementation, then the platforms against which the accelerator is advantageous or not, and finally the resource cost that distinguishes this work from the only comparable full implementation.

The two versions of the design presented in this work expose a frequency/cycle trade-off. Applying the Karatsuba algorithm to 32-bit multiplications shortens the critical path of the large-number multipliers and raises the maximum clock frequency from \SI{75}{\mega\hertz} to \SI{80}{\mega\hertz}, but it also increases the total cycle count of key-pair and signature generation (Tables \ref{tabla:compKG} and \ref{tabla:compSD}). Because the penalty of the additional clock cycles exceeds the gains from the higher frequency, the 64-bit version running at \SI{75}{\mega\hertz} remains faster end to end.

The comparison against general-purpose processors shows where the proposed implementation presents benefits. Against software that emulates floating point, the regime Peregrino targets, the accelerator reduces the signature-generation cycle count on every platform, including the ARM Cortex-A9, the emulated Intel run of \cite{Pornin2019}, and the ARM Cortex-M4 of \cite{Pornin2019, Pornin2025}. Meanwhile, processors with a native FPU compute key-pair and signature generation in fewer clock cycles than the accelerator, and processors clocked in the \si{\giga\hertz} range finish faster in absolute time even when they spend more cycles, since Peregrino runs at 75--\SI{80}{\mega\hertz}.

Regarding key-pair generation and verification the picture is mixed: Peregrino leads the ARM Cortex-M4 of \cite{Pornin2019} and the Microblaze, but not the emulated Intel run of \cite{Pornin2019} or the updated ARM Cortex-M4 of \cite{Pornin2025}.

The practical consequence is that the accelerator is most valuable on edge nodes that lack a native FPU, and that on more capable platforms it is best used selectively for the operations where it still leads, such as signature verification. Its distinguishing contribution is therefore completeness and low resource cost rather than raw speed.

That resource cost is what most clearly separates Peregrino from prior work. The
only other full hardware implementation of Falcon, the HLS-generated
FalconTakesOff \cite{FalconTakesOff}, reaches a lower cycle count but at a
resource utilization that Peregrino reduces by 1,9$\times$ in LUTs, 3,4$\times$ in
FFs, 2,7$\times$ in BRAMs, and 9,9$\times$ in DSPs (Table \ref{tabla:computil}).
Summed across its three parts, that design would occupy 118\%, 52\%, 33\%, and
191\% of the LUTs, FFs, BRAMs, and DSPs available on the
\textit{Xilinx XC7A200T-1SBG484C}. This implies that the amount of needed LUTs
and DSPs exceeds the capacity of one device, so
the complete HLS scheme requires at least two FPGAs of this kind. Peregrino fits the entire
scheme on one, the property that makes it deployable in the resource-constrained
edge setting this work targets.

It should be highlighted that Peregrino does not improve the area-time product (ATP) over the HLS design, but improves the feasibility of the implementation of the scheme. On the target class of device FalconTakesOff has an unbounded ATP because it does not fit. Below the capacity of a single mid-range FPGA device, resource fit is a binary constraint rather than a continuous cost, and Peregrino satisfies this restriction.

These results also mark the limitations of the current design. Peregrino runs at
75--\SI{80}{\mega\hertz}, so despite needing fewer clock cycles than several reference
platforms, its latency stays above that of \si{\giga\hertz}-class CPUs. Thus,
raising the operating frequency is the most direct path to closing that gap. On
platforms that expose a native FPU, key-pair and signature generation fall below
the accelerator's performance, which confines its clear advantage to FPU-less targets
and to signature verification. In addition, the overhead of data transfers derived from
the \textit{w/o-load} rows of Tables \ref{tabla:compSD} and \ref{tabla:compVV}
is bound to the MicroBlaze register interface, so a dedicated DMA path could recover most of it.

This work does not evaluate side-channel resistance. Peregrino targets a trusted edge node in which the accelerator and its host processor share a security boundary, and the design does not assume a physical attack model. The implemented modules differ in their exposure: NTRUGen and HashToPoint operate on data that is public or exercised only once per key pair, so timing variation in these blocks does not leak information with each signature. The \textit{ffSampling} module, by contrast, executes on every signature, and its variable latency depends on the secret Gaussian samples it draws, which is the same class of leakage that recovers Falcon's secret key from a single power trace \cite{karabulutdown}. Because Peregrino's modular architecture treats each block as a replaceable unit, the emulated floating-point multiplier at the core of the sampler could be substituted for a masked implementation such as the one reported in \cite{karabulutmasking}, at the 5,42--43,31$\times$ area cost the authors measure. We scope this substitution, and a full side-channel evaluation of the resulting design, as future work.

\section{Conclusion}\label{sec:conclusion}

This work presented Peregrino, the first hardware accelerator of the
complete Falcon digital signature scheme, i.e., key-pair generation, signature
generation, and signature verification, designed from scratch as
hand-written RTL. Every module was designed to be functionally equivalent
to the reference software, instantiated under a wrapper exposing an AXI
Lite interface to a MicroBlaze soft-core, and validated against the NIST KAT vectors.

The central result is that the presented design brings the complete
scheme within the resource budget of a single mid-range FPGA. The only
prior full implementation is generated by high-level synthesis and its
combined resource demand exceeds the capacity of one \textit{Xilinx
XC7A200T-1SBG484C}, while Peregrino fits the entire scheme on one such device,
using 1,9$\times$ fewer LUTs, 3,4$\times$ fewer FFs, 2,7$\times$ fewer
BRAMs, and 9,9$\times$ fewer DSPs. As a peripheral of an on-chip
MicroBlaze, it additionally reduces key-pair generation, signature
generation, and signature verification clock cycles by 92\%, 96\%, and
85\% over the emulated floating-point reference software.

Two directions follow from this design. Raising the accelerator's
75--\SI{80}{\mega\hertz} operating frequency and replacing the register-based host
interface with a dedicated DMA path would close most of the remaining
latency gap to high-performance processors. Independently, the modular
architecture allows the emulated floating-point multiplier at the core of
the sampler to be replaced with a masked implementation, which together
with a full side-channel evaluation is left as future work toward
deployment in untrusted settings.

\section*{Acknowledgment}

This work was supported by the INARTRANS 4.0 project PLEC2023-010343, funded by MICIU/AEI /10.13039/501100011033.

\ifCLASSOPTIONcaptionsoff
  \newpage
\fi

\printbibliography

\end{document}